\documentclass{article}
\usepackage{amssymb}
\usepackage{cite}

\input{tcilatex}
\begin{document}

\author{J. G. Cardoso\thanks{%
jorge.cardoso@udesc.br} \\
%EndAName
Department of Mathematics\\
Centre for Technological Sciences-UDESC\\
Joinville 89219-710 SC\\
Brazil.}
\title{ The non-singular Trautman-Kopczy\'{n}ski big-bang model and a
torsional spinor description of dark matter}
\date{ }
\maketitle

\begin{abstract}
A view is taken up whereby the non-singular Trautman-Kopczy\'{n}ski big-bang
creation of the Universe produced a highly torsional hot state at early
stages of the cosmic evolution which particularly brought about the
formation of a dark matter cloud. It is thus assumed that the combination of
Einstein-Cartan's theory with the torsionful version of the two-component $%
\varepsilon $-formalism of Infeld and van der Waerden supplies a natural
local description of dark matter in terms of uncharged spin-one massive
fields. In the case of either handedness, the pertinent spinor field
equation arises directly from a suitable form of the world Bianchi identity.
It appears that each such field equation bears a term that is thought of as
carrying part of the information on the mass of the fields. The whole
information turns out to be extracted by well prescribed derivatives of
certain couplings involving the fields and spinor torsion pieces in such a
way that the mass of dark matter is really thought of as arising from the
interaction between the fields and spinor torsion.
\end{abstract}

\section{Introduction}

It is well known that the theory of general relativity, as formulated in
terms of Einstein's field equations, gives rise to isotropic and homogeneous
cosmological models which predict the occurrence of singular gravitational
colapses in a way that does not depend upon both the physical contents of a
certain class of energy-momentum tensors and the symmetries of the models
[1-7]. According to this framework, there was at least one moment during the
evolution of the Universe at which the density of matter and energy was
infinite. Several attempts to circumvent this singularity situation were
made by just introducing a cosmological constant into the geometric side of
the generally relativistic field equations, but such a procedure
nevertheless has not been fully satisfactory from the theoretical viewpoint
[8,9].

In the work of Ref. [10], it was suggested for the first time that
Einstein-Cartan's theory [11,12] could be utilized for drawing up
alternative cosmological models which impede the production of singular
gravitational and cosmological collapses. Soon after the publication of this
work, it was shown [13] that Einstein-Cartan's equations admit a
two-parameter family of spherically symmetric solutions of the Friedmann
type, which supply a lower bound for the final radius of a gravitationally
collapsing cloud of dust, thereby providing a construction of what may be
surely designated as \textit{Trautman-Kopczy\'{n}ski cosmological models}.
More recently, the work of Ref. [14] has notably given a clear explanation
on how spacetime torsion at a microscopic level may generate a gravitational
repulsion that prevents the cosmological singularity to occur, while
presenting an explicit solution to the cosmological spatial flatness and
horizon problems by supplying physically plausible torsional mechanisms
without having to call upon any cosmic inflationary scenario.

An intrinsic property of Einstein-Cartan's theory relies upon the fact that
the characteristic asymmetry borne by the Ricci tensor for any torsionful
world affine connexion always entails the presence of asymmetric
energy-momentum tensors on the right-hand sides of the field equations. The
skew parts of such tensors were originally identified [15-17] with sources
for densities of intrinsic angular momentum of matter that presumably
generate spacetime torsion locally. Thus, as remarked in Ref. [10], spin
density of matter plays a physical role in Einstein-Cartan's theory which is
analogous to that played by mass in general relativity.

Amongst the most elementary spacetime properties of Einstein-Cartan's
theory, there is a striking feature associated to this theory which bears a
deeper mathematical character, namely, the fact that any spacetime endowed
with a torsionful affinity admits a local spinor structure in much the same
way as in the case of generally relativistic spacetimes [18,19]. This
admissibility has supported the construction of an essentially unique
torsional extension [20] of the famous two-component spinor $\gamma
\varepsilon $-formalisms of Infeld and van der Waerden for general
relativity [21,22]. Such as traditionally posed, the torsionless $%
\varepsilon $-formalism supplied the first description of gravitons [23]
wherein the relevant degrees of freedom are all carried by the totally
symmetric part of one of the Witten curvature spinors for a Riemann tensor
[24], with a systematic adaptation of this description to the torsionless $%
\gamma $-formalism having been exhibited later on in Ref. [22]. The work of
Ref. [22] also suggested a geometric definition of wave functions for the
cosmic microwave background, which show up in both torsionless formalisms as
spinors that occur in irreducible decompositions of Maxwell bivectors
produced by suitably contracted spin affinities. Within this torsionless
spinor framework, a complete set of wave equations that control the
spacetime propagation of the cosmic microwave background and gravitons, was
derived on the basis of the implementation of some formal algebraic
expansions and symbolic valence-reduction devices built up by Penrose (see
Ref. [12]).

The torsional extension referred to above of the classical Infeld-van der
Waerden formalisms, was primarily aimed at proposing a local description
whereby dark energy should be supposedly looked upon as a torsional cosmic
background [25]. Accordingly, dark energy fields were broadly defined as
uncharged spin-one massive fields which come from spin-affine pieces that
amount to gauge-invariant Proca potentials, and thence enter into
irreducible decompositions of torsional bivectors. Roughly speaking, these
invariant spin-affine contributions account for the underlying spacetime
torsionfulness, and likewise enter additively together with contracted
spin-affine pieces for the cosmic microwave background into overall
contracted spin affinities. It became evident that, in contradistinction to
the cosmic microwave background, dark energy can \textit{not} propagate
alone in spacetime. It was realized, in effect, that any torsional affine
potential must be accompanied by an adequate torsionless spin-affine
contribution such that the propagation of the dark energy background has to
be united together with that of the cosmic microwave background. However,
one background propagates in spacetime as if the other were absent whence
they do not interact with one another. Yet, the propagation of the cosmic
microwave background in regions of the Universe where the values of
torsional bivectors are negligible, may be described alone within the
torsionless framework. Moreover, the calculational procedures involved in
the derivation of the wave equations for the dark energy background, were
based upon the property that the algebraic expansions and valence-reduction
devices which had been utilized in the torsionless framework, still apply
formally to the torsionful framework.

In the present paper, we take up the view that the non-singular big-bang
creation of the Universe, as suggested by Trautman-Kopczy\'{n}ski
cosmological models [10,13], produced a highly torsional hot state at early
stages of the cosmic evolution which brought about the formation of the dark
energy background as well as the appearance of dark matter and gravitons. As
the theoretical framework we have been considering geometrically imposes
that the propagation of dark energy must be accompanied by the propagation
of the cosmic microwave background, it may be asserted that such physical
backgrounds were produced together. Hence, it can be said that we are
presumptively dealing here with a \textit{physical four-feature picture} of
the Universe which arises directly from the combination of Einstein-Cartan's
theory with a torsionful two-component spinor framework. It is our belief
that this combination should afford a transparent description of some of the
physical features of the Universe. By this point, we shall work on such
situation by proposing a description of dark matter within the context of
the torsionful $\varepsilon $-formalism. Dark matter is thus physically
characterized as uncharged spin-one massive fields which were produced
together with gravitons. This characterization stems only from the result
that wave functions for dark matter arise together with ones for gravitons
in the expansion for a curvature spinor of a torsionful spin affinity. We
stress that our attitude whereby dark matter should be theoretically
described by a curvature spinor, goes hand-in-hand with one of the basic
assumptions made in the original spinor description of gravitons [23]. We
will see that a spinor version of Einstein-Cartan's equations, which
naturally emerges in our framework, exhibits a physical identification as
dark matter of the densities of spinning matter that produce spacetime
torsion. So, as the world form of Einstein-Cartan's theory stands, it will
become manifest that it is only dark matter which constitutes the physical
source for spacetime torsion. We assume from the beginning that the Hubble
expansion of the Universe at accelerated stages of the cosmic evolution, is
partially due to the gravitational action on inner matter-energy contents of
a dense dark matter cloud. Our spinor field equations arise from a suitable
form of the world Bianchi identity, and suggest that the cosmic distribution
of dark matter generally did not possess spatial uniformity during the
evolutive eras of the Universe. As we believe, this may provide an easier
macroscopic explanation of the anisotropy of the presently observable
acceleration of the Universe (see, for instance, Refs. [26-28]). The
corresponding wave equations are derived out of utilizing the same
calculational techniques as those implemented for describing the dark energy
background. It will be seen by means of these wave equations that dark
matter interacts with gravitons and torsion, but the description of dark
energy as proposed in Ref. [25] will turn out to show that dark matter
interacts also with both of the cosmic backgrounds according to typical
coupling patterns. This insight seemingly identifies the main responsible
for the ocurrence of the observed angular anisotropy of the cosmic microwave
background [29]. It will appear that the dark energy background should
contribute significantly to the Hubble acceleration when it propagates
inside the dark matter cloud. Upon carrying out the pertinent derivation
procedures for the case of either handedness, we shall have to claim that
one of the terms borne by the respective field equation encodes part of the
information on the mass of dark matter. In the case of either handedness,
the whole information then turns out to be extracted by well prescribed
derivatives of couplings involving dark matter fields and spinor torsion
pieces in such a way that the mass of dark matter is really thought of as
arising from the interaction between the fields and spinor torsion.

Throughout the work, we will use the natural system of units in which $%
c=\hslash =1.$ All the world and spinor conventions adhered to in Ref. [25]
will be adopted in what follows. In particular, symmetrizations and
skew-symmetrizations over index blocks will be, respectively, denoted by
round and square brackets surrounding the indices singled out by the
symmetry operations, whilst vertical bars surrounding an index block will
mean that the indices sorted out are not to partake of a symmetry operation.

\section{\protect\LARGE Einstein-Cartan's Theory}

In the realm of Einstein-Cartan's theory, any spacetime carries a symmetric
metric tensor $g_{\mu \nu }$ along with a torsionful covariant derivative
operator $\nabla _{\mu }$ that satisfies the metric\footnote{%
It is convenient to attribute to $g_{\mu \nu }$ the local signature $(+---).$%
} compatibility condition $\nabla _{\mu }g_{\lambda \sigma }=0.$ The Riemann
tensor $R_{\mu \nu \lambda }{}^{\sigma }$ of $\nabla _{\mu }$ enters either
of the configurations%
\begin{equation}
D_{\mu \nu }u^{\alpha ...\beta }=R_{\mu \nu \tau }{}^{\alpha }u^{\tau
...\beta }{}+\cdots +R_{\mu \nu \tau }{}^{\beta }u^{\alpha ...\tau }{}
\label{e1}
\end{equation}%
and%
\begin{equation}
D_{\mu \nu }w_{\lambda ...\sigma }=-R_{\mu \nu \lambda }{}^{\tau }w_{\tau
...\sigma }-\cdots -R_{\mu \nu \sigma }{}^{\tau }w_{\lambda ...\tau },
\label{e2}
\end{equation}%
with $u^{\alpha ...\beta }{}$ and $w_{\lambda ...\sigma }$ being some world
tensors, and%
\begin{equation}
D_{\mu \nu }=2(\nabla _{\lbrack \mu }\nabla _{\nu ]}+T_{\mu \nu }{}^{\lambda
}\nabla _{\lambda }).  \label{e3}
\end{equation}%
The operator $D_{\mu \nu }$ thus possesses the Leibniz rule property. When
acting on a world-spin scalar $h,$ it gives $D_{\mu \nu }h=0$ by definition.
The object $T_{\mu \nu }{}^{\lambda }$ is the torsion tensor of $\nabla
_{\mu }.$ It amounts to the skew part $\Gamma _{\lbrack \mu \nu
]}{}^{\lambda }$ of the world affinity $\Gamma _{\mu \nu }{}^{\lambda }$ of $%
\nabla _{\mu },$ and thereby satisfies the defining property%
\begin{equation}
T_{\mu \nu }{}^{\lambda }=T_{[\mu \nu ]}{}^{\lambda },  \label{add1}
\end{equation}%
whence $D_{\mu \nu }=D_{[\mu \nu ]}.$ For the relevant Ricci tensor and
scalar, one has%
\begin{equation}
R_{\mu \nu }=R_{\mu \lambda \nu }{}^{\lambda },\text{ }R=R^{\lambda \sigma
}g_{\lambda \sigma }.  \label{e4}
\end{equation}

The tensor $R_{\mu \nu \lambda \sigma }{}$ possesses skewness in the indices
of the pairs $\mu \nu $ and $\lambda \sigma ,$ but the traditional
index-pair symmetry now fails to hold because of the applicability of the
cyclic identity\footnote{%
As posed in Ref. [12], the tensor $T_{\mu \lambda \sigma }{}$ conventionally
equals $(-2)$ times ours.}%
\begin{equation}
R_{[\mu \nu \lambda ]}{}^{\sigma }-2\nabla _{\lbrack \mu }T_{\nu \lambda
]}{}^{\sigma }+4T_{[\mu \nu }{}^{\tau }T_{\lambda ]\tau }{}^{\sigma }=0.
\label{add2}
\end{equation}%
Thus, the Ricci tensor of $\nabla _{\mu }$ bears asymmetry. Additionally,
the Bianchi identity should now read%
\begin{equation}
\nabla _{\lbrack \mu }R_{\nu \lambda ]\sigma }{}^{\rho }-2T_{[\mu \nu
}{}^{\tau }R_{\lambda ]\tau \sigma }{}^{\rho }=0.  \label{10}
\end{equation}%
By invoking the dualization schemes given in Ref. [12] and making some index
manipulations thereafter, we may rewrite (\ref{add2}) and (\ref{10}) in
terms of first-left duals as 
\begin{equation}
^{\ast }R^{\lambda }{}_{\mu \nu \lambda }+2\nabla ^{\lambda }{}^{\ast
}T_{\lambda \mu \nu }+4{}^{\ast }T_{\mu }{}^{\lambda \tau }{}T_{\lambda \tau
\nu }=0  \label{11}
\end{equation}%
and%
\begin{equation}
\nabla ^{\rho }{}^{\ast }R_{\rho \mu \lambda \sigma }+2{}^{\ast }T{}{}_{\mu
}{}^{\rho \tau }R_{\rho \tau \lambda \sigma }=0.  \label{12}
\end{equation}%
It should be clear from Eqs. (\ref{11}) and (\ref{12}) that both $^{\ast
}R^{\lambda }{}_{\mu \nu \lambda }$ and $\nabla ^{\rho }{}^{\ast }R_{\rho
\mu \lambda \sigma }$ do not vanish here in contraposition to the
torsionless framework.

The tensor $g_{\mu \nu }$ comes into play as a solution to Einstein-Cartan's
equations\footnote{%
The quantity $\kappa $ is identified with Einstein's gravitational constant
of general relativity.}%
\begin{equation}
R_{\mu \nu }-\frac{1}{2}Rg_{\mu \nu }=-\kappa E_{\mu \nu },  \label{e5}
\end{equation}%
whereas $T_{\mu \nu }{}^{\lambda }$ is locally related [12] to the spin
density of matter $S_{\mu \nu }{}^{\lambda }$ present in spacetime through%
\begin{equation}
T_{\mu \nu }{}^{\lambda }=-\kappa (S_{\mu \nu }{}^{\lambda }-S_{[\mu }g_{\nu
]}{}^{\lambda }),  \label{e6}
\end{equation}%
with $S_{\mu }\doteqdot S_{\mu \lambda }{}^{\lambda }.$ Equation (\ref{e6})
right away yields the trace relation%
\begin{equation}
T_{\mu }{}=\frac{\kappa }{2}S_{\mu },  \label{e7}
\end{equation}%
along with the definition $T_{\mu }\doteqdot T_{\mu \lambda }{}^{\lambda }.$
Consequently, we can write down the supplementary relationship%
\begin{equation}
-\kappa S_{\mu \nu }{}^{\lambda }=T_{\mu \nu }{}^{\lambda }-2T_{[\mu }g_{\nu
]}{}^{\lambda }.  \label{e8}
\end{equation}%
Hence, working out the relation (\ref{add2}) leads to the equation%
\begin{equation}
\nabla _{\lambda }T_{\mu \nu }{}^{\lambda }+2(\nabla _{\lbrack \mu }T_{\nu
]}{}+T_{\mu \nu }{}^{\lambda }T_{\lambda })=\kappa E_{[\mu \nu ]},
\label{e9Lin}
\end{equation}%
which amounts to the same thing as%
\begin{equation}
(\nabla _{\lambda }+\kappa S_{\lambda })S_{\mu \nu }{}^{\lambda }=-E_{[\mu
\nu ]}.  \label{e12}
\end{equation}%
We can thus say that the skew part of $E_{\mu \nu }$ is a source for $S_{\mu
\nu }{}^{\lambda },$ and thence also for $T_{\mu \nu }{}^{\lambda }.$

In fact, if the trace pattern%
\begin{equation}
T_{\mu }=\nabla _{\mu }\Phi  \label{add3}
\end{equation}%
is taken for granted, with $\Phi $ standing for a world-spin invariant, then
Eqs. (\ref{e8}) and (\ref{e12}) may be fitted together so as to yield the
statement%
\begin{equation}
\nabla _{\lambda }T_{\mu \nu }{}^{\lambda }=\kappa E_{[\mu \nu ]},
\label{add15}
\end{equation}%
which would also come straightaway from (\ref{e9Lin}). Whence, $T_{\mu \nu
}{}^{\lambda }$ appears to play a dynamical role similar to that played by $%
g_{\mu \nu },$ whilst\footnote{%
Making the choice (\ref{add3}) also implies that $\nabla _{\lbrack \mu
}S_{\nu ]}{}=\kappa S_{\mu \nu }{}^{\lambda }S_{\lambda }.$}%
\begin{equation}
\nabla _{\lbrack \mu }T_{\nu ]}{}+T_{\mu \nu }{}^{\lambda }T_{\lambda }=0
\label{add3Lin}
\end{equation}%
and%
\begin{equation}
\nabla _{\lambda }S_{\mu \nu }{}^{\lambda }+\nabla _{\lbrack \mu }S_{\nu
]}{}=-E_{[\mu \nu ]}.  \label{add3LinLin}
\end{equation}%
We shall see in Section 6 that the gradient model for $T_{\mu }$ as given by
(\ref{add3}), imparts a somewhat simpler form to our expression for the mass
of dark matter.

\section{Curvature Spinors and Derivatives}

It is shown in Refs. [20,25] that, in either formalism, the curvature
spinors for $\nabla _{\mu }$ occur in the prescriptions 
\begin{equation}
\check{D}_{AB}\zeta ^{C}=\varpi _{ABM}{}^{C}\zeta ^{M},\text{ }\check{D}%
_{A^{\prime }B^{\prime }}\zeta ^{C}=\varpi _{A^{\prime }B^{\prime
}M}{}^{C}\zeta ^{M},  \label{40}
\end{equation}%
where $\zeta ^{A}$ is a spin vector. In the $\varepsilon $-formalism, $%
\check{D}_{AB}$ and $\check{D}_{A^{\prime }B^{\prime }}$ enter the operator
bivector decomposition%
\begin{equation}
\Sigma _{AA^{\prime }}^{\mu }\Sigma _{BB^{\prime }}^{\nu }D_{\mu \nu
}=\varepsilon _{A^{\prime }B^{\prime }}\check{D}_{AB}+\varepsilon _{AB}%
\check{D}_{A^{\prime }B^{\prime }},  \label{34}
\end{equation}%
with $D_{\mu \nu }$ being given by Eq. (\ref{e3}) and the $\Sigma $-objects
amounting to appropriate connecting objects. The $\varepsilon $-objects are
the only covariant metric spinors for the formalism being allowed for since
they bear invariance under the action of the generalized Weyl gauge group
[20]. Both of them and any $\Sigma $-connecting object are usually taken as
covariantly constant entities such that, for instance,%
\begin{equation}
\nabla _{\mu }\varepsilon _{AB}=0.  \label{e34}
\end{equation}%
Indeed, any curvature spinors obey a general symmetry relation like%
\begin{equation}
\varpi _{ABCD}{}=\varpi _{(AB)CD}{},\text{ }\varpi _{A^{\prime }B^{\prime
}CD}{}=\varpi _{(A^{\prime }B^{\prime })CD}{}.  \label{36}
\end{equation}

It may be established [20] that the spinor pair $(\varpi _{AB(CD)}{},\varpi
_{A^{\prime }B^{\prime }(CD)}{})$ constitutes the irreducible decomposition
of the Riemann tensor for $\nabla _{\mu }$. We have, in effect, the
expression%
\begin{equation}
R_{AA^{\prime }BB^{\prime }CC^{\prime }DD^{\prime }}=\hspace{-1pt}%
(\varepsilon _{A^{\prime }B^{\prime }}\varepsilon _{C^{\prime }D^{\prime
}}\varpi _{AB(CD)}\hspace{-1pt}+\varepsilon _{AB}\varepsilon _{C^{\prime
}D^{\prime }}\varpi _{A^{\prime }B^{\prime }(CD)})+\text{c.c.},  \label{108}
\end{equation}%
together with its first-left dual%
\begin{equation}
^{\ast }R_{AA^{\prime }BB^{\prime }CC^{\prime }DD^{\prime }}=[-i(\varepsilon
_{A^{\prime }B^{\prime }}\varepsilon _{C^{\prime }D^{\prime }}\varpi
_{AB(CD)}\hspace{-1pt}-\varepsilon _{AB}\varepsilon _{C^{\prime }D^{\prime
}}\varpi _{A^{\prime }B^{\prime }(CD)})]+\text{c.c.},  \label{111}
\end{equation}%
where the symbol "c.c." denotes here as elsewhere an overall complex
conjugate piece. The unprimed curvature spinor is expandable as\footnote{%
From now onwards, we will for convenience employ the definitions $\varpi
_{AB(CD)}{}\doteqdot $ X$_{ABCD}$ and $\varpi _{A^{\prime }B^{\prime
}(CD)}\doteqdot \Xi _{A^{\prime }B^{\prime }CD}{}.$ We should also stress
that $\varepsilon _{A(C}\varepsilon _{D)B}=\varepsilon _{(A\mid
(C}\varepsilon _{D)\mid B)}.$}%
\begin{equation}
\text{X}_{ABCD}\hspace{-0.07cm}=\hspace{-0.07cm}\Psi _{ABCD}-\varepsilon
_{(A\mid (C}\xi _{D)\mid B)}-\frac{1}{3}\varkappa \varepsilon
_{A(C}\varepsilon _{D)B},  \label{41}
\end{equation}%
with%
\begin{equation}
\hspace{-0.07cm}\Psi _{ABCD}=\text{X}_{(ABCD)}\hspace{-0.07cm},\text{ }\xi
_{AB}=\text{X}^{M}{}_{(AB)M},\text{ }\varkappa =\text{X}_{LM}{}^{LM}
\label{42}
\end{equation}%
and $\Psi _{ABCD}$ defining a formal prototype of a wave function for
gravitons, which vanishes identically in case the underlying spacetime bears
conformal flatness. The quantity $\varkappa $ thus appears as a
complex-valued world-spin invariant that satisfies%
\begin{equation}
R=4\func{Re}\varkappa ,\text{ }^{\ast }R{}_{\mu \nu }{}^{\mu \nu }=4\func{Im}%
\varkappa .  \label{add15Lin}
\end{equation}%
We emphasize that the occurrence of the $\xi $-spinor in the expansion (\ref%
{41}), is just associated to the torsional property $R_{\mu \nu \lambda
\sigma }\neq R_{\lambda \sigma \mu \nu }.$ Then, the only symmetries borne
by the X-spinor are exhibited by%
\begin{equation}
\text{X}_{ABCD}=\text{X}_{(AB)(CD)}.  \label{add5}
\end{equation}

The contracted pieces $(\varpi _{ABM}{}^{M},\varpi _{A^{\prime }B^{\prime
}M}{}^{M})$ fulfill additivity relations that carry wave functions for the
cosmic microwave and dark energy backgrounds of both handednesses, namely
(for further details, see Ref. [25])%
\begin{equation}
\varpi _{ABM}{}^{M}=-2i(\phi _{AB}+\psi _{AB})  \label{43}
\end{equation}%
and%
\begin{equation}
\varpi _{A^{\prime }B^{\prime }M}{}^{M}=-2i(\psi _{A^{\prime }B^{\prime
}}+\phi _{A^{\prime }B^{\prime }}).  \label{44}
\end{equation}%
We can therefore recast the prescriptions (\ref{40}) into the form%
\begin{equation}
\check{D}_{AB}\zeta ^{C}=\text{X}_{ABM}{}^{C}\zeta ^{M}-i(\phi _{AB}+\psi
_{AB})\zeta ^{C}  \label{45}
\end{equation}%
and 
\begin{equation}
\check{D}_{A^{\prime }B^{\prime }}\zeta ^{C}=\Xi _{A^{\prime }B^{\prime
}M}{}^{C}\zeta ^{M}-i(\phi _{A^{\prime }B^{\prime }}+\psi _{A^{\prime
}B^{\prime }})\zeta ^{C}.  \label{46}
\end{equation}%
The prescriptions for computing $\check{D}$-derivatives of a covariant spin
vector $\eta _{A}$ can be immediately obtained from (\ref{45}) and (\ref{46}%
) by requiring that%
\begin{equation}
\check{D}_{AB}(\zeta ^{C}\eta _{C})=0,\text{ }\check{D}_{A^{\prime
}B^{\prime }}(\zeta ^{C}\eta _{C})=0,  \label{47}
\end{equation}%
and carrying out Leibniz expansions thereof.\footnote{%
It should be obvious that the Leibniz-rule property of $D_{\mu \nu }$ is set
forth to both $\check{D}_{AB}$ and $\check{D}_{A^{\prime }B^{\prime }}.$} We
thus obtain%
\begin{equation}
\check{D}_{AB}\eta _{C}=-\hspace{1pt}[\text{X}_{ABC}{}^{M}\eta _{M}-i(\phi
_{AB}+\psi _{AB})\eta _{C}]  \label{48}
\end{equation}%
and%
\begin{equation}
\check{D}_{A^{\prime }B^{\prime }}\eta _{C}=-\hspace{1pt}[\Xi _{A^{\prime
}B^{\prime }C}{}^{M}{}\eta _{M}-i(\phi _{A^{\prime }B^{\prime }}+\psi
_{A^{\prime }B^{\prime }})\eta _{C}],  \label{49}
\end{equation}%
along with the complex conjugates of Eqs. (\ref{45}) through (\ref{49}).
Hence, whenever $\check{D}$-derivatives of Hermitian spin tensors are
actually computed, all occurrent $\phi \psi $-contributions get cancelled.
Obviously, such a cancellation likewise happens when we allow the $\check{D}$%
-operators to act freely upon spin tensors having the same numbers of
covariant and contravariant indices of the same kind. In any such case, the
relevant expansion thus carries only gravitational contributions.

The development of Section 6 formally demands the use of derivatives of
geometric spinor densities [22, 25], which will be very briefly touched upon
now. The $\check{D}$-derivatives of a complex spin-scalar density $\alpha $
of weight $w$ are written as\footnote{%
The complex conjugate of the spin density $\alpha $ is said to possess an 
\textit{antiweight} $w$ such that $\mid \alpha \mid ^{2}$ possesses an 
\textit{absolute weight} $2w.$}%
\begin{equation}
\check{D}_{AB}\alpha =2iw\alpha (\phi _{AB}+\psi _{AB}),\text{ }\check{D}%
_{A^{\prime }B^{\prime }}\alpha =2iw\alpha (\phi _{A^{\prime }B^{\prime
}}+\psi _{A^{\prime }B^{\prime }}),  \label{50}
\end{equation}%
whence the patterns for $\check{D}$-derivatives of some spin-tensor density
can always be specified by symbolic expansions like 
\begin{equation}
\check{D}_{AB}(\alpha \Upsilon _{C...D})=(\check{D}_{AB}\alpha )\Upsilon
_{C...D}+\alpha \check{D}_{AB}\Upsilon _{C...D},  \label{53}
\end{equation}%
with $\Upsilon _{C...D}$ being a spin tensor. When $w<0$, a cancellation of
the $\phi \psi $-pieces will occur in the expansion (\ref{53}) as well if $%
\Upsilon _{C...D}$ is taken to carry $-2w$ indices and $\func{Im}\alpha \neq
0$ everywhere. A similar property also holds for situations that involve
outer products between contravariant spin tensors and complex spin-scalar
densities having suitable positive weights. In the $\varepsilon $-formalism,
each of the objects $\phi _{AB},$ $\psi _{AB}$ and $\xi _{AB}$ neatly fits
in with the situations just described, inasmuch as each of them is
effectively identified thereabout with a spin-tensor density of weight $-1.$

Equations (\ref{e3}) and (\ref{34}) give rise to the derivative operators%
\begin{equation}
\check{D}_{AB}=\Delta _{AB}+2\tau _{AB}{}^{\mu }\nabla _{\mu },\text{ }%
\Delta _{AB}\doteqdot -\nabla _{(A}^{C^{\prime }}\nabla _{B)C^{\prime }},
\label{35}
\end{equation}%
where $\tau _{AB}{}^{\mu }$ is borne by the bivector expansion for $%
T_{\lambda \sigma }{}^{\mu },$ that is to say,%
\begin{equation}
T_{AA^{\prime }BB^{\prime }}{}^{\mu }=\varepsilon _{A^{\prime }B^{\prime
}}\tau _{AB}{}^{\mu }+\text{c.c.}.  \label{17}
\end{equation}%
In conjunction with (\ref{45}) and (\ref{46}), we thus have the differential
relations%
\begin{equation}
\Delta _{AB}\zeta ^{C}=\check{D}_{AB}\zeta ^{C}-2\tau _{AB}{}^{\mu }\nabla
_{\mu }\zeta ^{C}  \label{add18}
\end{equation}%
and%
\begin{equation}
\Delta _{A^{\prime }B^{\prime }}\zeta ^{C}=\check{D}_{A^{\prime }B^{\prime
}}\zeta ^{C}-2\tau _{A^{\prime }B^{\prime }}{}^{\mu }\nabla _{\mu }\zeta
^{C},  \label{add23}
\end{equation}%
along with the ones for $\eta _{C}.$ The defining property $D_{\mu \nu }h=0$
can then be recovered as%
\begin{equation}
\Delta _{AB}h=-2\tau _{AB}{}^{\mu }\nabla _{\mu }h.  \label{52Lin}
\end{equation}%
Hence, we can spell out the expansion%
\begin{equation}
\nabla _{AC^{\prime }}\nabla _{B}^{C^{\prime }}=\Delta _{AB}+\frac{1}{2}%
\varepsilon _{AB}\square ,  \label{56}
\end{equation}%
as well as its contravariant version, with $\square =\nabla _{\mu }\nabla
^{\mu }.$

\section{Einstein-Cartan's Theory in Spinor Form}

Equations (\ref{108}) and (\ref{41}) supply us with the following expression
for the spinor version of $R_{\mu \nu }$%
\begin{equation}
R_{AA^{\prime }BB^{\prime }}=\varepsilon _{AB}\varepsilon _{A^{\prime
}B^{\prime }}\func{Re}\varkappa -[(\varepsilon _{A^{\prime }B^{\prime }}\xi
_{AB}+\Xi _{A^{\prime }B^{\prime }AB})+\text{c.c.}].  \label{add17}
\end{equation}%
Under certain affine circumstances, the symmetric part of Einstein-Cartan's
equations leads to the limiting case of general relativity, but this is not
of a primary interest at this stage. Instead, we should now allow for the
skew part%
\begin{equation}
R_{[\mu \nu ]}=-\kappa E_{[\mu \nu ]},  \label{add19}
\end{equation}%
whose spinor version is, then, constituted by%
\begin{equation}
\varepsilon _{A^{\prime }B^{\prime }}\xi _{AB}+\text{c.c.}=\kappa
(\varepsilon _{A^{\prime }B^{\prime }}\check{E}_{AB}+\text{c.c.}).
\label{add21}
\end{equation}%
Hence, if we implement a decomposition for the spin density of matter $%
S_{\mu \nu }{}^{\lambda }$ like the one exhibited by (\ref{17}), namely,%
\begin{equation}
S_{AA^{\prime }BB^{\prime }}{}^{\mu }=\varepsilon _{A^{\prime }B^{\prime }}%
\check{S}_{AB}{}^{\mu }+\text{c.c.},  \label{e30}
\end{equation}%
after calling for Eqs. (\ref{e6}) and (\ref{e12}), we will get the relation%
\begin{equation}
\tau _{AB}{}^{CC^{\prime }}=-\kappa (\check{S}_{AB}{}^{CC^{\prime }}+\frac{1%
}{2}S_{(A}^{C^{\prime }}\varepsilon _{B)}{}^{C}),  \label{add50}
\end{equation}%
together with%
\begin{equation}
(\nabla _{\mu }+\kappa S_{\mu })\check{S}_{AB}{}^{\mu }=-\frac{1}{\kappa }%
\xi _{AB}.  \label{e31}
\end{equation}%
Equation (\ref{e8}) thus gets translated into%
\begin{equation}
-\kappa \check{S}_{AB}{}^{CC^{\prime }}=\tau _{AB}{}^{CC^{\prime
}}+T_{(A}^{C^{\prime }}\varepsilon _{B)}{}^{C}.  \label{add51}
\end{equation}

The dynamical role played by $T_{\mu \nu }{}^{\lambda }$ becomes
considerably enhanced when the spinor version of (\ref{add15}) is set up. We
have, in effect,%
\begin{equation}
\nabla _{\mu }\tau _{AB}{}^{\mu }=\xi _{AB},  \label{e32}
\end{equation}%
while the choice (\ref{add3LinLin}) should be transcribed as%
\begin{equation}
\nabla _{\mu }\check{S}_{AB}{}^{\mu }+\frac{1}{2}\nabla _{C^{\prime
}(A}S_{B)}^{C^{\prime }}=-\frac{1}{\kappa }\xi _{AB}.  \label{add5Lin}
\end{equation}%
A glance at Eqs. (\ref{e9Lin}), (\ref{add21}) and (\ref{e31}) tells us that,
as far as the physical inner structure of Einstein-Cartan's theory is
concerned, a $\xi $-curvature spinor must be taken as the only source for
spacetime torsion and densities of intrinsic angular momentum of matter.%
\footnote{%
This statement remains valid even when the pattern (\ref{add3}) is brought
forth along with Eq. (\ref{e32}).} We will elaborate a little further upon
this point in the concluding Section.

\section{Field Equations}

The attitude we have been taking herein involves identifying the world field
equation for dark matter as the statement%
\begin{equation}
\nabla ^{\rho \text{ }\ast }R^{\lambda }{}_{[\rho \sigma ]\lambda }=-2^{\ast
}T^{\lambda \rho \tau }R_{\tau \lbrack \rho \sigma ]\lambda },  \label{1}
\end{equation}%
which is nothing else but a skew contracted version of the Bianchi identity (%
\ref{12}). The spinor version of $^{\ast }R^{\lambda }{}_{[\rho \sigma
]\lambda }$ yields the expansion%
\begin{equation}
\Sigma _{BB^{\prime }}^{\rho }\Sigma _{CC^{\prime }}^{\sigma }{}^{\ast
}R^{\lambda }{}_{[\rho \sigma ]\lambda }={}^{\ast }R^{DD^{\prime
}}{}_{(BC)[B^{\prime }C^{\prime }]DD^{\prime }}+\text{c.c.},  \label{2}
\end{equation}%
which is written out explicitly as%
\begin{equation}
^{\ast }R^{DD^{\prime }}{}_{(BC)[B^{\prime }C^{\prime }]DD^{\prime }}+\text{%
c.c.}=-i(\varepsilon _{B^{\prime }C^{\prime }}\xi _{BC}{}-\text{c.c.}).
\label{3}
\end{equation}%
Thus, the left-hand side of Eq. (\ref{1}) takes the form%
\begin{equation}
\nabla ^{BB^{\prime }}[-i(\varepsilon _{B^{\prime }C^{\prime }}\xi _{BC}{}-%
\text{c.c.})]=-i(\nabla _{C^{\prime }}^{B}\xi _{BC}{}-\text{c.c.}),
\label{4}
\end{equation}%
whence the entries of the pair $(\xi _{AB},\xi _{A^{\prime }B^{\prime }})$
assumably constitute wave functions of opposite handednesses for dark
matter, with each of which thus possessing six real independent components.
For the right-hand side of Eq. (\ref{1}), we have the correspondence%
\footnote{%
For $^{\ast }T^{\lambda \rho \tau },$ we have the expansion $i(\varepsilon
^{DB}\tau ^{D^{\prime }B^{\prime }AA^{\prime }}-$ c.c.$).$}%
\begin{equation}
-2^{\ast }T^{\lambda \rho \tau }R_{\tau \lbrack \rho \sigma ]\lambda
}\longmapsto -i(\varepsilon ^{DB}\tau ^{D^{\prime }B^{\prime }AA^{\prime }}-%
\text{c.c.})(\varepsilon _{B^{\prime }C^{\prime }}R_{AA^{\prime
}(BC)L^{\prime }}{}^{L^{\prime }}{}_{DD^{\prime }}+\text{c.c.}),  \label{5}
\end{equation}%
together with the expression%
\begin{eqnarray}
R_{AA^{\prime }(BC)L^{\prime }}{}^{L^{\prime }}{}_{DD^{\prime }}
&=&-\varepsilon _{A^{\prime }D^{\prime }}\text{X}_{A(BC)D}{}+\Xi _{A^{\prime
}D^{\prime }D(C}\varepsilon _{B)A}  \nonumber \\
&&+\varepsilon _{A(B}\varepsilon _{C)D}\text{X}_{L^{\prime }A^{\prime
}D^{\prime }}{}^{L^{\prime }}+\varepsilon _{D(C}\Xi _{B)AA^{\prime
}D^{\prime }}.  \label{6}
\end{eqnarray}

Towards carrying through the derivation of our field equations, it is
convenient to require the unprimed and primed wave functions to bear
algebraic independence throughout spacetime. This requirement enables us to
rearrange some of the complex conjugate pieces of Eqs. (\ref{5}) and (\ref{6}%
) to the extent that the left-right handednesses of the fields become
transparently separated. We should then expand the whole Riemann term of (%
\ref{5}) in agreement with%
\begin{eqnarray}
2R_{\tau \lbrack \rho \sigma ]\lambda } &\longmapsto &[\varepsilon
_{B^{\prime }C^{\prime }}(-\varepsilon _{A^{\prime }D^{\prime }}\text{X}%
_{A(BC)D}{}+\varepsilon _{D(C}\Xi _{B)AA^{\prime }D^{\prime }})  \nonumber \\
&&+\varepsilon _{BC}(\Xi _{ADD^{\prime }(C^{\prime }}\varepsilon _{B^{\prime
})A^{\prime }}+\varepsilon _{A^{\prime }(B^{\prime }}\varepsilon _{C^{\prime
})D^{\prime }}\text{X}_{LAD}{}^{L})]+\text{c.c.},  \label{7}
\end{eqnarray}%
whose unprimed X-contributions, by virtue of Eq. (\ref{41}), are expressed as%
\begin{equation}
\text{X}_{A(BC)D}=\Psi _{ABCD}-\frac{1}{2}\varepsilon _{AD}\xi _{BC}+\frac{1%
}{6}\varkappa \varepsilon _{A(B}\varepsilon _{C)D}  \label{8}
\end{equation}%
and%
\begin{equation}
\text{X}_{LAD}{}^{L}=-(\xi _{AD}-\frac{1}{2}\varkappa \varepsilon _{AD}).
\label{9}
\end{equation}%
Therefore, putting into effect the rearrangement given by (\ref{7}) allows
us to work out naively the $\tau R$-couplings that occur in Eq. (\ref{5}).

For the kernel $\tau ^{D^{\prime }B^{\prime }AA^{\prime }},$ we thus utilize
trivial index-displacement rules for obtaining the X-contribution\footnote{%
We should notice that one of the couplings which occur on the left-hand side
of (\ref{15}) annihilates the $\Psi $-term of (\ref{8}).}%
\begin{eqnarray}
&&\varepsilon ^{DB}\tau ^{D^{\prime }B^{\prime }AA^{\prime }}(-\varepsilon
_{B^{\prime }C^{\prime }}\varepsilon _{A^{\prime }D^{\prime }}\text{X}%
_{A(BC)D}+\varepsilon _{BC}\varepsilon _{A^{\prime }(B^{\prime }}\varepsilon
_{C^{\prime })D^{\prime }}\text{X}_{LAD}{}^{L})  \nonumber \\
&=&\tau _{A^{\prime }C^{\prime }}{}^{AA^{\prime }}\xi _{AC}-\frac{1}{2}%
\varkappa \tau _{A^{\prime }C^{\prime }C}{}^{A^{\prime }},  \label{15}
\end{eqnarray}%
along with the $\Xi $-one%
\begin{eqnarray}
&&\varepsilon ^{DB}\tau ^{D^{\prime }B^{\prime }AA^{\prime }}(\varepsilon
_{B^{\prime }C^{\prime }}\varepsilon _{D(C}\Xi _{B)AA^{\prime }D^{\prime
}}+\varepsilon _{BC}\Xi _{ADD^{\prime }(C^{\prime }}\varepsilon _{B^{\prime
})A^{\prime }})  \nonumber \\
&=&2\tau ^{B^{\prime }D^{\prime }B}{}_{C^{\prime }}\Xi _{BCB^{\prime
}D^{\prime }}-\tau ^{B^{\prime }D^{\prime }B}{}_{D^{\prime }}\Xi
_{BCB^{\prime }C^{\prime }}.  \label{19}
\end{eqnarray}%
In turn, for the kernel $\tau ^{DBAA^{\prime }},$ we have%
\begin{eqnarray}
&&\varepsilon ^{D^{\prime }B^{\prime }}\tau ^{DBAA^{\prime }}(-\varepsilon
_{B^{\prime }C^{\prime }}\varepsilon _{A^{\prime }D^{\prime }}\text{X}%
_{A(BC)D}+\varepsilon _{BC}\varepsilon _{A^{\prime }(B^{\prime }}\varepsilon
_{C^{\prime })D^{\prime }}\text{X}_{LAD}{}^{L})  \nonumber \\
&=&\tau ^{ABD}{}_{C^{\prime }}\Psi _{ABCD}+\frac{3}{2}\tau
^{AB}{}_{CC^{\prime }}\xi _{AB}-2\tau ^{AB}{}_{BC^{\prime }}\xi _{AC}-\frac{5%
}{6}\varkappa \tau ^{A}{}_{CAC^{\prime }}  \label{21}
\end{eqnarray}%
and%
\begin{eqnarray}
&&\varepsilon ^{D^{\prime }B^{\prime }}\tau ^{DBAA^{\prime }}(\varepsilon
_{B^{\prime }C^{\prime }}\varepsilon _{D(C}\Xi _{B)AA^{\prime }D^{\prime
}}+\varepsilon _{BC}\Xi _{ADD^{\prime }(C^{\prime }}\varepsilon _{B^{\prime
})A^{\prime }})  \nonumber \\
&=&\tau ^{AB}{}_{B}{}^{A^{\prime }}\Xi _{ACA^{\prime }C^{\prime }}-\tau
^{AB}{}_{C}{}^{A^{\prime }}\Xi _{ABA^{\prime }C^{\prime }}.  \label{23}
\end{eqnarray}%
It follows that, if we take into consideration the relation%
\begin{equation}
\tau ^{AB}{}_{BB^{\prime }}-\tau _{A^{\prime }B^{\prime }}{}^{AA^{\prime
}}=T_{B^{\prime }}^{A}=\Sigma _{B^{\prime }}^{\mu A}T_{\mu },  \label{26}
\end{equation}%
after performing some further index manipulations and implementing the
required algebraic independence between the conjugate wave functions, we
will arrive at the field equation%
\begin{equation}
\nabla _{C^{\prime }}^{B}\xi _{BC}{}+m_{CC^{\prime }}=\sigma _{CC^{\prime
}}^{(\text{X})}+\sigma _{CC^{\prime }}^{(\Xi )},  \label{e15}
\end{equation}%
together with 
\begin{equation}
m_{CC^{\prime }}=(T_{C^{\prime }}^{A}-3\tau ^{AB}{}_{BC^{\prime }})\xi _{AC}+%
\frac{3}{2}\tau ^{AB}{}_{CC^{\prime }}{}\xi _{AB}  \label{e16}
\end{equation}%
and the complex conjugates of (\ref{e15}) and (\ref{e16}). The $\sigma $%
-pieces of Eq. (\ref{e15}) are complex sources for $\xi _{BC},$ which are
irreducibly given by%
\begin{equation}
\sigma _{CC^{\prime }}^{(\text{X})}=-\tau ^{ABD}{}_{C^{\prime }}\Psi
_{ABCD}+\varkappa (\frac{1}{2}T_{CC^{\prime }}+\frac{1}{3}\tau
^{A}{}_{CAC^{\prime }})  \label{e17}
\end{equation}%
and%
\begin{equation}
\sigma _{CC^{\prime }}^{(\Xi )}=\tau ^{AB}{}_{C}{}^{A^{\prime }}\Xi
_{ABA^{\prime }C^{\prime }}+2\tau ^{A^{\prime }B^{\prime }A}{}_{C^{\prime
}}\Xi _{ACA^{\prime }B^{\prime }}-T^{AA^{\prime }}\Xi _{CAA^{\prime
}C^{\prime }}.  \label{e18}
\end{equation}

We can see that the $m$-terms borne by the field equations for $\xi _{AB}$
and $\xi _{A^{\prime }B^{\prime }},$ arise strictly from the gravitational
interaction between torsion kernels and the fields themselves, while the
corresponding sources carry only couplings of torsion kernels with $%
\varkappa \Psi \Xi $-curvatures. In Section 7, a world interpretation of
this feature will be made.

\section{Wave Equations}

To derive the wave equations that govern the propagation in spacetime of the 
$\xi $-fields, we follow up the procedure which amounts to implementing the
expansions and derivatives exhibited previously to work out the $\xi $-field
equation for either handedness. For $\xi _{BC},$ say, we thus let the
operator $\nabla _{A}^{C^{\prime }}$ act on both sides of Eq. (\ref{e15}).
By making use of the contravariant form of (\ref{56}), we then obtain%
\footnote{%
It has been unnecessary here to stagger the indices of any symmetric
two-index configuration.}%
\begin{equation}
\frac{1}{2}\square \xi {}_{AC}+\check{D}_{A}^{B}\xi _{BC}-(2\tau _{A}^{B\mu
}\nabla _{\mu }\xi _{BC}+\nabla _{A}^{C^{\prime }}m_{CC^{\prime }})=-(\nabla
_{A}^{C^{\prime }}\sigma _{CC^{\prime }}^{(\text{X})}+\nabla _{A}^{C^{\prime
}}\sigma _{CC^{\prime }}^{(\Xi )}),  \label{67}
\end{equation}%
with Eq. (\ref{35}) having been employed. As $\xi _{BC}$ stands for a
two-index spin-tensor density of weight $-1,$ the derivative $\check{D}%
_{A}^{B}\xi _{BC}$ consists of a purely gravitational expansion, which may
be made up of%
\begin{equation}
\check{D}_{(A}^{B}\xi _{C)B}=\frac{2}{3}\varkappa \xi _{AC}-\Psi
_{AC}{}^{BD}\xi _{BD}  \label{51}
\end{equation}%
and%
\begin{equation}
\check{D}_{[A}^{B}\xi _{C]B}=\varepsilon _{AC}\xi ^{BD}\xi _{BD}.  \label{52}
\end{equation}

It is obvious that the symmetry of $\xi _{AC}$ brings about the occurrence
of the relation%
\begin{equation}
\nabla _{\mu }m^{\mu }-\Delta ^{BD}\xi _{BD}=\nabla ^{\mu }\sigma _{\mu }^{(%
\text{X})}+\nabla ^{\mu }\sigma _{\mu }^{(\Xi )},  \label{73}
\end{equation}%
which just amounts to the skew part in $A$ and $C$ of the statement (\ref{67}%
). Hence, accounting for (\ref{52}) and (\ref{73}), we get the subsidiary
condition%
\begin{equation}
2(\xi ^{BD}\xi _{BD}+\tau ^{BD\mu }\nabla _{\mu }\xi _{BD})+\nabla _{\mu
}m^{\mu }-(\nabla ^{\mu }\sigma _{\mu }^{(\text{X})}+\nabla ^{\mu }\sigma
_{\mu }^{(\Xi )})=0.  \label{75}
\end{equation}%
The overall symmetric piece of (\ref{67}) thus reads%
\begin{equation}
\frac{1}{2}\square \xi {}_{AC}+\check{D}_{(A}^{B}\xi _{C)B}-2(\nabla _{\mu
}\xi _{B(A})\tau _{C)}^{B\mu }-\nabla _{(A}^{C^{\prime }}m_{C)C^{\prime
}}=-(\nabla _{(A}^{C^{\prime }}\sigma _{C)C^{\prime }}^{(\text{X})}+\nabla
_{(A}^{C^{\prime }}\sigma _{C)C^{\prime }}^{(\Xi )}).  \label{e19}
\end{equation}%
With the help of Eq. (\ref{e32}), we can still reexpress the explicit $%
\nabla \xi \tau $-term of (\ref{e19}) as%
\begin{equation}
(\nabla _{\mu }\xi _{B(A})\tau _{C)}^{B\mu }=\nabla _{\mu }(\xi _{B(A}\tau
_{C)}^{B\mu }),  \label{61}
\end{equation}%
provided that $\xi _{(A}^{B}\xi _{C)B}\equiv 0.$

We saw that Einstein-Cartan's theory assigns a dynamical meaning to torsion
kernels. Based upon this fact and the formal commonness between the
couplings borne by $\nabla _{(A}^{C^{\prime }}m_{C)C^{\prime }}$ and that
shown up by Eq. (\ref{61}), we claim that the mass $\mu _{DM}$ of dark
matter is produced by the interaction between the $\xi $-fields and torsion
kernels, in accordance with the Klein-Gordon prescription\footnote{%
We recall that the natural system of units has been adopted by us from the
beginning.}%
\begin{equation}
\frac{1}{2}\mu _{DM}^{2}\xi _{AC}=-2\nabla _{\mu }(\xi _{B(A}\tau
_{C)}^{B\mu })-\nabla _{(A}^{C^{\prime }}m_{C)C^{\prime }}  \label{e73}
\end{equation}%
and the reality requirement $\mu _{DM}^{2}>0.$ The wave equation for $\xi
_{AC}$ comes about when (\ref{51}) and (\ref{e73}) are inserted into (\ref%
{e19}). We thus end up with%
\begin{equation}
(\square +\mu _{DM}^{2}+\frac{4}{3}\varkappa )\xi _{AB}-2\Psi
_{AB}{}^{CD}\xi _{CD}=-2(\nabla _{(A}^{C^{\prime }}\sigma _{B)C^{\prime }}^{(%
\text{X})}+\nabla _{(A}^{C^{\prime }}\sigma _{B)C^{\prime }}^{(\Xi )}),
\label{73Lin}
\end{equation}%
along with the complex conjugate of (\ref{73Lin}).

\section{World Dynamics, Mass Terms and Energy Density}

Within the geometric context of Section 2, we can deduce the equivalent
relations [20]%
\begin{equation}
\widetilde{\nabla }_{\mu }g_{\lambda \sigma }-2T_{\mu (\lambda \sigma
)}=0\Leftrightarrow \Gamma _{\mu }=\widetilde{\Gamma }_{\mu }+T_{\mu
}=\partial _{\mu }\log (-\mathfrak{g})^{1/2},  \label{n2}
\end{equation}%
where%
\begin{equation}
\widetilde{\Gamma }_{\mu }=\widetilde{\Gamma }_{\mu \lambda }{}^{\lambda },%
\text{ }\Gamma _{(\mu \lambda )\sigma }\doteqdot \widetilde{\Gamma }_{\mu
\lambda \sigma },  \label{n9}
\end{equation}%
with $\widetilde{\nabla }_{\mu }$ amounting to the covariant derivative
operator for $\widetilde{\Gamma }_{\mu \lambda \sigma },$ and $\mathfrak{g}$
denoting the determinant of $g_{\mu \nu }.$ Allowing for the secondary
metric condition%
\begin{equation}
\widetilde{\nabla }_{\lambda }g_{\mu \nu }=0,  \label{addLin1}
\end{equation}%
thus implies that $T_{\mu }=0$ everywhere in spacetime. Noticeably, this
implication yields the property%
\begin{equation}
T_{\mu \nu \lambda }=T_{[\mu \nu \lambda ]},  \label{in2}
\end{equation}%
such that $T_{\mu \nu \lambda }$ would possess only four real independent
components if the condition (\ref{addLin1}) were actually implemented.
Therefore, once we are given a solution to Eq. (\ref{e5}), with the
components of $E_{[\mu \nu ]}$ being accordingly prescribed at the outset,
we may promptly make use of the contortion tensor for $T_{\mu \nu \lambda }$
and identify $\widetilde{\Gamma }_{\mu \lambda \sigma }$ with a Christoffel
connexion, not only to determine all the components of $\Gamma _{(\mu
\lambda )\sigma },$ but also to select out a set of outer products that obey
the relationship%
\begin{equation}
2(\Sigma _{(\mu }^{00^{\prime }}\Sigma _{\nu )}^{11^{\prime }}-\Sigma _{(\mu
}^{01^{\prime }}\overline{\Sigma _{\nu )}^{01^{\prime }}})=g_{\mu \nu }.
\label{n1}
\end{equation}%
Consequently, if the expressions (\ref{add17}) and (\ref{add19}) were
accounted for afterwards, we would become able to set the corresponding
components of $R_{[\mu \nu ]},$ $\xi _{AC}$ and $\xi _{A^{\prime }C^{\prime
}}.$ Hence, in view of (\ref{addLin1}), we could utilize the simplified
expansion%
\begin{equation}
\frac{1}{\sqrt{-\mathfrak{g}}}\partial _{\lambda }(\sqrt{-\mathfrak{g}}%
T_{\mu \nu }{}^{\lambda })+2\widetilde{\Gamma }_{\lambda \lbrack \mu
}{}^{\sigma }T_{\nu ]\sigma }{}^{\lambda }=\kappa E_{[\mu \nu ]},  \label{n}
\end{equation}%
to integrate out Eq. (\ref{add15}) towards getting the components of $\tau
_{AB}{}^{\mu }$ and making up the information on those of $\Gamma _{\mu \nu
\lambda }{}$ which would, in turn, provide us with the ones of%
\begin{equation}
R_{\mu \nu \lambda }{}^{\rho }\doteqdot 2(\partial _{\lbrack \mu }\Gamma
_{\nu ]\lambda }{}^{\rho }+\Gamma _{\lbrack \mu \mid \tau \mid }{}^{\rho
}\Gamma _{\nu ]\lambda }{}^{\tau })  \label{nLin}
\end{equation}%
and%
\begin{equation}
^{\ast }R{}_{\mu \nu \lambda \sigma }=\frac{1}{2}\sqrt{-\mathfrak{g}}e_{\mu
\nu \rho \tau }R^{\rho \tau }{}_{\lambda \sigma },  \label{nLinLin}
\end{equation}%
with $e_{\lambda \sigma \rho \tau }$ being one of the invariant Levi-Civitta
world densities. In this way, all the components that enter the
prescriptions (\ref{e16}) and (\ref{e73}) might be determined consistently,
and an estimation of $\mu _{DM}^{2}$ could then be attained. It should be
emphasized at this point that the transcription (\ref{add5Lin}) guarantees
the legitimacy of the strong assumption on spin alignment, which is rooted
in the obtainment of Kopczy\'{n}ski solutions [13] (see also Ref. [11]).

In order to formulate the world dynamics of dark matter in a more explicit
manner, it may be expedient to define\footnote{%
There is no relationship between $B_{\mu }$ and any contracted spin
affinity, contrarily to the case of the cosmic backgrounds.}%
\begin{equation}
f_{\mu \nu }=2\nabla _{\lbrack \mu }B_{\nu ]},  \label{n3}
\end{equation}%
where $B_{\lambda }$ is a dark matter potential, and%
\begin{equation}
f_{\mu \nu }\doteqdot \text{ }^{\ast }R^{\lambda }{}_{[\mu \nu ]\lambda }.
\label{n4}
\end{equation}%
The action for dark matter is settled as%
\begin{equation}
S_{DM}=\int_{\omega }\sqrt{-\mathfrak{g}}(-\frac{1}{4}f^{\mu \nu }f_{\mu \nu
}+\frac{1}{2}M^{\mu }B_{\mu }+j^{\mu }B_{\mu })d^{4}x,  \label{n5}
\end{equation}%
where $\omega $ denotes a spacetime volume whose closure is compact, and%
\begin{equation}
d^{4}x=\frac{1}{4!}e_{\lambda \sigma \rho \tau }dx^{\lambda }\wedge
dx^{\sigma }\wedge dx^{\rho }\wedge dx^{\tau }.  \label{n5Lin}
\end{equation}%
In setting up Eq. (\ref{n5}), we have split the right-hand side of the
statement (\ref{1}) as follows%
\begin{equation}
2^{\ast }T^{\lambda \rho \tau }R_{\tau \lbrack \rho \mu ]\lambda }=M_{\mu
}+j_{\mu }.  \label{n7}
\end{equation}%
The only reason for this splitting procedure is, indeed, that the derivation
of the field equations of Section 5 formally brings forward a disjunction
between sources and torsion-field couplings, to which $S_{DM}$ must be
subject. We will see in a moment that, whenever $T_{\mu }=0,$ the
contributions $M_{\mu }$ and $j_{\mu }$ must carry the information on the
remaining pieces borne by the $m$-pattern and $\sigma $-sources of Eq. (\ref%
{e15}), respectively (see Eqs. (\ref{n50}) and (\ref{n90}) below).

The least-action principle involving $S_{DM}$ reads off%
\begin{equation}
\delta S_{DM}=0.  \label{n6}
\end{equation}%
By definition, the $\delta $-variation bears linearity and enjoys the
Leibniz-rule property, in addition to being defined so as to commute with
partial derivatives and integrations. The quantity $\delta B_{\mu }$ takes
arbitrary values in $\omega $ and vanishes on the boundary $\partial \omega $
of $\omega ,$ while $\delta j_{\mu }=0$ on the closure of $\omega .$ We
suppose that $\delta M_{\mu }$ should be expressed as $\mu _{m}^{2}\delta
B_{\mu },$ with $\mu _{m}^{2}$ being identified with the contribution to $%
\mu _{DM}^{2}$ that comes from the $m$-term exhibited as Eq. (\ref{e16}),
whence%
\begin{equation}
\delta (M^{\mu }B_{\mu })=2\mu _{m}^{2}B^{\mu }\delta B_{\mu }=2M^{\mu
}\delta B_{\mu }.  \label{n100}
\end{equation}%
Since%
\begin{equation}
\frac{1}{4}\delta (f^{\mu \nu }f_{\mu \nu })=\frac{1}{2}f^{\mu \nu }\delta
f_{\mu \nu }=f^{\mu \nu }(\partial _{\mu }\delta B_{\nu }-T_{\mu \nu
}{}^{\lambda }\delta B_{\lambda }),  \label{n6Lin}
\end{equation}%
after performing the integration by parts%
\begin{eqnarray}
&&\int_{\omega }\sqrt{-\mathfrak{g}}f^{\mu \nu }\partial _{\mu }\delta
B_{\nu }d^{4}x  \nonumber \\
&=&\int_{\partial \omega }\sqrt{-\mathfrak{g}}f^{\mu \nu }\delta B_{\nu
}d^{3}x_{\mu }-\int_{\omega }[\frac{1}{\sqrt{-\mathfrak{g}}}\partial _{\mu }(%
\sqrt{-\mathfrak{g}}f^{\mu \nu })]\delta B_{\nu }\sqrt{-\mathfrak{g}}d^{4}x,
\label{n9Lin}
\end{eqnarray}%
with%
\begin{equation}
d^{3}x_{\mu }=\frac{1}{3!}e_{\mu \lambda \sigma \rho }dx^{\lambda }\wedge
dx^{\sigma }\wedge dx^{\rho },  \label{n10}
\end{equation}%
and fitting pieces together, we thus obtain the equations of motion%
\begin{equation}
\frac{1}{\sqrt{-\mathfrak{g}}}\partial _{\mu }(\sqrt{-\mathfrak{g}}f^{\mu
\lambda })+f^{\mu \nu }T_{\mu \nu }{}^{\lambda }+M^{\lambda }=-j^{\lambda }.
\label{n15}
\end{equation}%
Now, if we take into account the relations (\ref{n2}), (\ref{addLin1}) and (%
\ref{n7}) together with the expansion%
\begin{eqnarray}
\nabla _{\mu }f^{\mu \nu } &=&\partial _{\mu }f^{\mu \nu }+\Gamma _{\mu
\lambda }{}^{\mu }f^{\lambda \nu }+\Gamma _{\mu \lambda }{}^{\nu }f^{\mu
\lambda }  \nonumber \\
&=&\partial _{\mu }f^{\mu \nu }+(\widetilde{\Gamma }_{\mu }{}-T_{\mu
})f^{\mu \nu }+T_{\mu \lambda }{}^{\nu }f^{\mu \lambda }  \nonumber \\
&=&\frac{1}{\sqrt{-\mathfrak{g}}}\partial _{\mu }(\sqrt{-\mathfrak{g}}f^{\mu
\nu })-2T_{\mu }f^{\mu \nu }+T_{\mu \lambda }{}^{\nu }f^{\mu \lambda },
\label{n15Lin}
\end{eqnarray}%
effectively setting $T_{\mu }=0$ in it, we will recover the field equation (%
\ref{1}) as%
\begin{equation}
\nabla _{\mu }f^{\mu \lambda }+M^{\lambda }=-j^{\lambda },  \label{n50}
\end{equation}%
while obtaining the wave equation%
\begin{equation}
\square B_{\mu }+R_{\mu }{}^{\lambda }B_{\lambda }-\nabla _{\mu }(\nabla
^{\lambda }B_{\lambda })-2T_{\mu }{}^{\lambda \sigma }\nabla _{\sigma
}B_{\lambda }=-2^{\ast }T^{\lambda \rho \tau }R_{\tau \lbrack \rho \mu
]\lambda }.  \label{n16}
\end{equation}%
It becomes clear that, if $T_{\mu }=0$ then the constituents of the
splitting (\ref{n7}) should occur in the correspondences%
\begin{equation}
2\func{Im}\Sigma _{\mu }^{CC^{\prime }}m_{CC^{\prime }}=M_{\mu },\text{ }2%
\func{Im}\Sigma _{\mu }^{CC^{\prime }}(\sigma _{CC^{\prime }}^{(\text{X}%
)}+\sigma _{CC^{\prime }}^{(\Xi )})=-j_{\mu }.  \label{n90}
\end{equation}

From Eq. (\ref{n5}), it follows that the overall Lagrangian density for dark
matter may be symbolically written as%
\begin{equation}
\mathfrak{L}_{DM}=\mathfrak{L}_{M}+\mathfrak{L}_{\sigma },  \label{n12Lin}
\end{equation}%
with the individual pieces%
\begin{equation}
\mathfrak{L}_{M}=-\frac{1}{4}f^{\mu \nu }f_{\mu \nu }+\frac{1}{2}M^{\mu
}B_{\mu },\text{ }\mathfrak{L}_{\sigma }=j^{\mu }B_{\mu }.  \label{n12}
\end{equation}%
Hence, by employing the following definition for the respective
energy-momentum tensor%
\begin{equation}
T_{\mu \nu }^{DM}=g_{\mu \nu }\mathfrak{L}_{M}-f_{\mu \lambda }g_{\rho \nu }%
\frac{\partial \mathfrak{L}_{M}}{\partial \nabla _{\rho }B_{\lambda }},
\label{n70}
\end{equation}%
we get the explicit expression\footnote{%
We should note that $f_{[\mu \mid \lambda \mid }f_{\nu ]}{}^{\lambda }=0$
such that $T_{\mu \nu }^{DM}$ still bears symmetry.}%
\begin{equation}
T_{\mu \nu }^{DM}=g_{\mu \nu }(-\frac{1}{4}f^{\lambda \sigma }f_{\lambda
\sigma }+\frac{1}{2}M^{\lambda }B_{\lambda })+f_{\mu \lambda }f_{\nu
}{}^{\lambda }.  \label{n71}
\end{equation}%
Evidently, rather than allowing for Eq. (\ref{n16}) to obtain the components
of $B_{\mu },$ we may absorb the property (\ref{in2}) to integrate the
equation%
\begin{equation}
\partial _{\lbrack \mu }B_{\nu ]}-T_{\mu \nu }{}^{\lambda }B_{\lambda }=%
\frac{1}{2}f_{\mu \nu }.  \label{n73}
\end{equation}%
With the components of $M_{\mu }$ at hand, such as given by Eq. (\ref{n90}),
we can say that the above procedure would easily produce the value%
\begin{equation}
\mu _{m}=\sqrt{\frac{M^{\lambda }B_{\lambda }}{B^{\tau }B_{\tau }}}.
\label{n74}
\end{equation}%
Equation (\ref{n50}) could, then, be reinstated as%
\begin{equation}
\nabla _{\mu }f^{\mu \lambda }+\mu _{m}^{2}B^{\lambda }=-j^{\lambda },
\label{n75}
\end{equation}%
whilst the energy density carried by $f_{\mu \nu }$ would be expressed by%
\begin{equation}
T_{00}^{DM}=g_{00}(-\frac{1}{4}f^{\lambda \sigma }f_{\lambda \sigma }+\frac{1%
}{2}\mu _{m}^{2}B^{\lambda }B_{\lambda
})+f_{01}f_{0}{}^{1}+f_{02}f_{0}{}^{2}+f_{03}f_{0}{}^{3}.  \label{n76}
\end{equation}

\section{Concluding Remarks and Outlook}

The expansion (\ref{41}) is what suggests that gravitons and dark matter
were produced together by the big-bang creation of the Universe, whence we
can say that the earliest states of very high density of spin matter must
have occurred in the absence of conformal flatness. Nevertheless, the spinor
form of Einstein-Cartan's theory shows us that it is solely dark matter
which produces spacetime torsion, but the theory of dark matter we have
proposed particularly asserts that the sources for dark matter must be
prescribed in terms of well defined couplings between curvatures and
torsion. Hence, our theory appears to assign a double physical character to
dark matter. According to the formulation of Sections 5 and 6, the
propagation of dark matter in spacetime is controlled by the wave equation (%
\ref{73Lin}) and its complex conjugate. A systematic treatment of these
differential equations can be carried out by utilizing a torsional extension
of the methods provided by Ref. [30]. Such a treatment would thus supply a
description of the effects of curvature and torsion on the wave functions
that are presumably ascribed to dark matter, while possibly shedding some
light on the assumption that the massiveness of dark matter arises strictly
from the interaction between dark matter fields and torsion kernels just as
simply prescribed by Eq. (\ref{e73}).

It was seen that the implementation of the gradient pattern (\ref{add3})
ensures the formal commonness of $\nabla _{(A}^{C^{\prime }}m_{C)C^{\prime
}} $ with the right-hand side of Eq. (\ref{61}), whence we may think of the
mass of dark matter as being expressed as%
\begin{equation}
\mu _{DM}=\sqrt{\frac{[-4\nabla _{\mu }(\xi _{B(A}\tau _{C)}^{B\mu
})-2\nabla _{(A}^{C^{\prime }}m_{C)C^{\prime }}]\xi ^{AC}}{\xi _{AC}\xi ^{AC}%
}}.  \label{N1}
\end{equation}%
The significance of the derivation procedures exibited in Section 6
partially originates from the fact that Eq. (\ref{73Lin}) brings out the
overall information on the mass of dark matter, while the field equation (%
\ref{e15}) can supply only the part of this information that is carried by
the prescription (\ref{e16}). Remarkably enough, this observation still
applies to the description of the cosmic microwave and dark energy
backgrounds we had referred to in Section 1. Whenever the cosmic backgrounds
propagate inside dark matter, the microwave background thus acquires a
certain amount of mass $\mu _{CMB}$ while the dark energy background
acquires an effective mass $\mu _{eff}.$ The corresponding relations are
formally the same as the one given by (\ref{e73}), namely,%
\begin{equation}
\mu _{CMB}^{2}\phi _{AC}=-4\nabla _{\mu }(\phi _{B(A}\tau _{C)}^{B}{}^{\mu
})-2\nabla _{(A}^{C^{\prime }}M_{C)C^{\prime }}  \label{N2}
\end{equation}%
and%
\begin{equation}
\mu _{DE}^{2}\psi _{AC}=-4\nabla _{\mu }(\psi _{B(A}\tau _{C)}^{B}{}^{\mu
})-2\nabla _{(A}^{C^{\prime }}\mathcal{M}_{C)C^{\prime }},  \label{N3}
\end{equation}%
with the characteristic expressions%
\begin{equation}
M_{CC^{\prime }}=2(T_{C^{\prime }}^{A}\phi _{AC}-\tau ^{AB}{}_{CC^{\prime
}}\phi _{AB}),\text{ }\mathcal{M}_{CC^{\prime }}=2(T_{C^{\prime }}^{A}\psi
_{AC}-\tau ^{AB}{}_{CC^{\prime }}\psi _{AB})  \label{N4}
\end{equation}%
and the requirements%
\begin{equation}
\mu _{CMB}^{2}>0,\text{ }\mu _{DE}^{2}>0.  \label{N5}
\end{equation}%
We may conclude that any wave functions for dark energy and dark matter can
not be taken as null fields as they propagate in spacetime, whereas any one
for the cosmic microwave background must not be taken as a null field when
it propagates in torsional regions of the Universe. In the $\varepsilon $%
-formalism, the terms of the wave equations that keep track of the
propagation of the cosmic backgrounds can therefore be rearranged as%
\begin{equation}
(\square +\mu _{CMB}^{2}+\frac{4}{3}\varkappa )\phi _{AB}-2\Psi
{}_{AB}{}^{CD}\phi _{CD}+2\xi _{(A}^{C}\phi _{B)C}=0  \label{N6}
\end{equation}%
and%
\begin{equation}
(\square +\mu _{eff}^{2}+\frac{4}{3}\varkappa )\psi _{AB}-2\Psi
{}_{AB}{}^{CD}\psi _{CD}+2\xi _{(A}^{C}\psi _{B)C}=2m^{2}\tau _{AB}{}^{\mu
}{}A_{\mu },  \label{N7}
\end{equation}%
where $A_{\mu }$ is a gauge-invariant affine potential, $m$ amounts to the
Proca mass of $\psi _{AB}$ and 
\begin{equation}
\mu _{eff}=\sqrt{m^{2}+\mu _{DE}^{2}}.  \label{N8}
\end{equation}

It is worth pointing out that the procedures involved in the derivation of
Eq. (\ref{73Lin}), yield the occurrence of a $\xi ^{2}$-term at the
intermediate steps of the pertinent calculations, which is eliminated from
the basis of the theory because of Eq. (\ref{75}) and the symmetry of $\xi
_{AB}.$ Our wave equations for dark matter of both handednesses thus bear
linearity in contrast to the ones for gravitons that occur in general
relativity, which carry couplings of the type $\Psi _{MN(AB}{}\Psi
{}_{CD)}{}^{MN}.$ It is an observational fact that curvature fields possess
darkness, but dark matter nonetheless interacts with spin $1/2$ fermions
through couplings which look like $\xi _{A}^{B}\psi _{B}$ and $\xi
_{B^{\prime }}^{A^{\prime }}\chi ^{B^{\prime }}$ while gravitons do not. A
noteworthy feature of our framework is related to its prediction that both
the cosmic backgrounds interact with dark matter, but the relevant couplings
are stringently borne by the wave equations for the backgrounds. Such a
peculiar occurrence had taken place in connection with the derivation of the
wave equations for the microwave background and gravitons in torsionless
spacetime environments. It could take place once again if a torsional
description of gravitons were carried out within the framework we have been
allowing for, since any $\varepsilon $-formalism wave functions for
gravitons $(\Psi {}^{ABCD}{},\Psi {}_{ABCD}{})$ are four-index spin-tensor
densities of weight $\pm 2.$ Hence, the dark energy background must
contribute significantly to the Hubble acceleration when it propagates
inside the dark matter cloud. The cosmic distribution of dark matter as
coming from Eqs. (\ref{e15}) and (\ref{73Lin}) is far from bearing
uniformity, whence we can say that the couplings $\Psi {}_{AB}{}^{CD}\phi
_{CD}$ and $\xi _{(A}^{C}\phi _{B)C}$ may supply an explanation of the
observed angular anisotropy of the cosmic microwave background in
non-conformally flat regions of the Universe. Of course, a similar
explanation can also be accomplished from the couplings that occur in the
wave equation for $\psi _{AB}$ such that we could expect that the dark
energy background bears anisotropy. Then, notwithstanding the result that
the attainment of solutions to the wave equations (\ref{N6}) and (\ref{N7})
would likewise lead to theoretical values for the angular anisotropies of
the cosmic microwave and dark energy backgrounds, both of these anisotropies
could have been predicted beforehand in a qualitative way.

Of the utmost importance as regards the physical characters of the work just
presented is, in fact, the achievement of a geometric estimation of the mass
of dark matter along with an observational confirmation that the cosmic
microwave background should behave as a massive field as it propagates
inside dark matter. The work of Section 7 describes a world situation which
enables one to estimate the contribution to the mass of dark matter that
comes from the pattern (\ref{e16}) along with the contribution owing to the
derivative (\ref{61}). It turns out that if the condition (\ref{addLin1}) is
used up in the case of a conformally flat Kopczy\'{n}ski solution, then all
wave functions for gravitons should be set equal to zero together with the
torsion trace $T_{\mu },$ and we might settle the world version of Eq. (\ref%
{73Lin}) as the dark matter statement%
\begin{equation}
\square f_{\mu \nu }+M_{\mu \nu }^{\ast }+F_{\mu \nu }^{\ast }=s_{\mu \nu
}^{\ast },  \label{N500}
\end{equation}%
with the bivector formulae%
\begin{equation}
M_{\mu \nu }^{\ast }=2\func{Im}(\mu _{DM}^{2}\Sigma _{\mu B^{\prime
}}^{A}\Sigma _{\nu }^{BB^{\prime }}\xi _{AB}),\text{ }F_{\mu \nu }^{\ast }=%
\frac{8}{3}\func{Im}(\varkappa \Sigma _{\mu B^{\prime }}^{A}\Sigma _{\nu
}^{BB^{\prime }}\xi _{AB})  \label{N501}
\end{equation}%
and%
\begin{equation}
s_{\mu \nu }^{\ast }=-4\func{Im}[\Sigma _{\mu B^{\prime }}^{A}\Sigma _{\nu
}^{BB^{\prime }}(\nabla _{(A}^{C^{\prime }}\hat{s}_{B)C^{\prime }}^{(\text{X}%
)}+\nabla _{(A}^{C^{\prime }}\hat{s}_{B)C^{\prime }}^{(\Xi )})],
\label{N502}
\end{equation}%
whence we could equally well estimate the Klein-Gordon masses $\mu _{CMB}$
and $\mu _{DE}$ too, in conformity with the Trautman-Kopczy\'{n}ski
cosmological models. Having at our disposal an estimated value of $\mu _{DE}$
obtained in this way, would permit us to draw a comparison of it with those
which come from the negative-pressure interpretation of dark energy as
provided in detail by the work of Ref. [31]. This might eventually yield an
interesting estimation of the Proca mass of $\psi _{AB}.$ It would be
worthwhile, also, to compare the values of acquired masses coming from
covariant derivatives of couplings between fields and torsion kernels with
the ones that arise from the dynamical description of the very early stages
of the evolution of the Universe as formulated by the theory of cosmological
pertubations [32, 33]. This latter comparison may perhaps make it possible
to investigate whether dark matter and dark energy should play some
important role in the establishment of a definitive inflationary solution to
the cosmological spatial flatness and horizon problems. Thus, one could
compare the results that come from the popular scalar-field scenario with
those supplied by the dark dynamics of%
\begin{equation}
\mathfrak{L}_{D}=\mathfrak{L}_{EC}+\mathfrak{L}_{M}+\mathfrak{L}_{E},
\label{N9}
\end{equation}%
where $\mathfrak{L}_{EC}$ stands for the Einstein-Cartan Lagrangian density
whilst $\mathfrak{L}_{M}$ is given by Eq. (\ref{n12}). The contribution $%
\mathfrak{L}_{E}$ amounts to the Lagrangian density for dark energy, which
is expressed as%
\begin{equation}
\mathfrak{L}_{E}=-\frac{1}{4}F^{\mu \nu }F_{\mu \nu }+\frac{1}{2}E^{\mu
}A_{\mu }+\frac{1}{2}m^{2}A_{\mu }A^{\mu },  \label{N12}
\end{equation}%
with $E^{\mu }$ being defined in a way similar to $M^{\mu },$ and%
\begin{equation}
F_{\mu \nu }\doteqdot 2(\nabla _{\lbrack \mu }A_{\nu ]}+T_{\mu \nu
}{}^{\lambda }A_{\lambda })  \label{N15}
\end{equation}%
amounting to a bivector that arises from a suitably contracted torsional
spin-affinity, as utilized in Ref. [25].

Cosmological descriptions in spacetimes equipped with torsionful affinities,
have supplied many very striking physical features. Among these, the most
important ones are associated to the occurrence of geometric wave functions
in spinor descriptions, which could not emerge within any purely world
framework. The explanations concerning the cosmological singularity
prevention and gravitational repulsion, which come straightforwardly from
the world form of Einstein-Cartan's theory, do not require at all the use of
the $\gamma \varepsilon $-formalisms. However, as we said before, it is our
belief that either torsional spinor formalism recovers in terms of
predictions a bosonic four-feature picture of the Universe which involves
the cosmic backgrounds, gravitons and dark matter, that is to say,%
\[
\{\phi _{AB},\psi _{AB},\Psi _{ABCD},\xi _{AB}\}. 
\]

ACKNOWLEDGEMENT: I should acknowledge a reviewer for making some situational
remarks that have produced a considerable improvement upon the paper
exhibited here.

\end{document}